\documentclass[twocolumn,prl,amsmath,amssymb,floatfix]{revtex4-1}
\usepackage{amsmath}
\usepackage{amssymb}
\usepackage{graphicx}
\usepackage{bm}
\usepackage{color}
\usepackage{hyperref}
\usepackage{graphicx}
\usepackage{upgreek}
\usepackage{scalerel}
\usepackage{multirow}
\usepackage{pgffor}
\usepackage{pdfpages}
\usepackage{mathdots}
\usepackage{float}
\usepackage{silence}
\usepackage{physics} 
\usepackage{lscape}   
\usepackage{color, colortbl}
\usepackage[table]{xcolor}
\usepackage{comment}  
\makeatletter
\AtBeginDocument{\let\LS@rot\@undefined}
\makeatother

\begin{document}

\title{Distinguishing trivial and topological zero energy states in long  nanowire junctions}
\author{Jorge Cayao}
\affiliation{Department of Physics and Astronomy, Uppsala University, Box 516, S-751 20 Uppsala, Sweden}
\author{Annica M. Black-Schaffer}
\affiliation{Department of Physics and Astronomy, Uppsala University, Box 516, S-751 20 Uppsala, Sweden}

\date{\today}
\begin{abstract}
The emergence of zero energy states in non-topological superconductors represents an inevitable problem that  obscures the proper identification of zero energy Majorana bound states (MBSs) and prevents their use as topologically protected qubits.  In this Research Letter we investigate  long superconductor-normal-superconductor junctions where  trivial zero energy states, robust over a large range of parameters, appear as a result of helicity  and confinement in the normal region.  We demonstrate that both equilibrium supercurrents and critical currents are sensitive to variations in the length of the superconductor regions in the topological phase hosting MBSs, but, remarkably, no such length dependence exists when robust, but trivial, zero energy states are present. This strikingly different response originates from the non-local nature of the MBSs and we, therefore, propose it as a simple protocol for distinguishing between trivial and topological zero energy states.
\end{abstract}
\maketitle

The realization of topological superconductivity and Majorana bound states (MBSs) have generated a great deal of attention in physics, not only because they represent a new superconducting state but also due to their potential for applications \cite{Aguadoreview17,lutchyn2018majorana,zhang2019next,beenakker2019search,doi:10.1146/annurev-conmatphys-031218-013618,2020Aguado}.  Among the most striking properties of MBSs is their nonlocality and associated non-Abelian statistics, crucial for applications in topological quantum computation  \cite{kitaev,PhysRevLett.86.268,RevModPhys.80.1083,Sarma:16,alicea2011non,PhysRevX.6.031016}.  This has led to a series of experiments using nanowires with spin-orbit coupling (SOC) and proximity-induced $s$-wave superconductivity, where a large enough external magnetic field is predicted to drive the system into a topological  phase with MBSs at the wire end points \cite{PhysRevLett.105.077001,PhysRevLett.105.177002}.   So far, the primary detection scheme of MBSs has been a zero-bias conductance peak of height $2e^{2}/h$ at zero temperature \cite{PhysRevLett.98.237002,PhysRevLett.103.237001,PhysRevB.82.180516} due to their unique charge neutral nature, with several recent experimental results in apparent agreement \cite{Mourik:S12,Higginbotham,Deng16,Albrecht16,zhang16,Suominen17,Nichele17,gul2018ballistic}.

Recent studies have reported, however, that the observed zero-bias  peaks might instead have a mundane trivial origin and hence do not prove the presence of MBSs, as originally thought. In fact, it has been shown that Andreev bound states (ABSs) can emerge at zero energy for magnetic fields well below the topological transition \cite{PhysRevB.86.100503,PhysRevB.86.180503,PhysRevB.91.024514,JorgeEPs,StickDas17,Ptok2017Controlling,Fer18,PhysRevB.98.245407,PhysRevLett.123.107703,PhysRevB.100.155429,PhysRevLett.123.217003,10.21468/SciPostPhys.7.5.061,avila2019non,PhysRevResearch.2.013377,PhysRevLett.125.017701,PhysRevLett.125.116803,Olesia2020,yu2020non,valentini2020nontopological,prada2019andreev}, with properties similar to those of MBSs. While some such trivial zero energy states (ZESs) are easily removable by varying system parameters, other trivial ZESs remain robust for a large range of parameters, making the identification of MBSs extremely challenging.

\begin{figure}[!t]
	\centering
	\includegraphics[width=.49\textwidth]{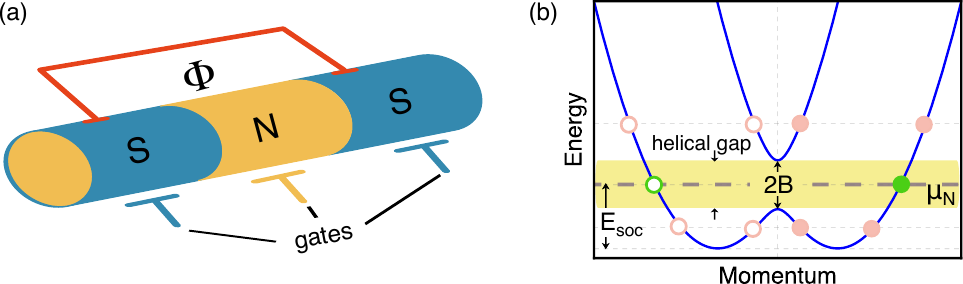}
	\caption{(a) Semiconducting nanowire with SOC covered by $s$-wave superconductors (S, blue) leaving an uncovered central region  in the normal (N, yellow) state. S regions are connected by a grounded superconducting loop allowing control of the phase difference $\phi$ via the flux $\Phi$. Gate voltages individually control the chemical potential in each region. (b) Energy dispersion in the normal state with an applied perpendicular magnetic field producing a gap of size $2B$ (yellow) and centered around $\mu_{\rm N}=0$. Inside this gap the system exhibits only two Fermi points, representing counterpropagating states with different spins (solid and open green circles), and this is thus a \textit{helical gap}. For $\mu$ above or below this gap there are four Fermi points (pink circles). Visibility of the helical gap is lost when $B\gg E_{\rm SOC}$, with $E_{\rm SOC}$ being the SOC energy.}
	\label{Fig1}
\end{figure}

An experimentally relevant regime with highly robust but trivial ZESs is found in superconductor-normal-superconductor (SNS) junctions with long N regions, due to  confinement and helicity in the normal region N \cite{PhysRevB.86.180503,PhysRevB.91.024514,PhysRevB.91.024514,JorgeEPs}.  We stress that typical experimental nanowire setups are likely to belong to this type of junction, as confinement effects appear when contacting a conductor between  e.g.~S leads, as in Fig.\,\ref{Fig1}(a), and also when gating the N region \cite{PhysRevB.98.085125}. Helicity is present whenever the N region is gated such that its chemical potential lies within the so-called helical energy gap, opened due to a combined effect of SOC and magnetic field; see Fig.\,\ref{Fig1}(b). 
This helical regime is found at low magnetic fields and thus occurs as an inevitable interim phase before the topological phase at higher fields, which further complicates detecting MBSs. Despite recent efforts \cite{PhysRevB.93.245404,PhysRevB.96.195307,PhysRevB.97.045421,schrade2018andreev,PhysRevB.97.165302,PhysRevB.97.214502,PhysRevB.97.161401,PhysRevB.99.155159,PhysRevB.100.241408,PhysRevLett.123.117001,PhysRevB.101.014512,PhysRevB.102.045111,PhysRevB.102.045303,doi:10.1002/qute.201900110,pan2020threeterminal,liu2020topological,PhysRevLett.124.036801,ricco2020topological,thamm2020transmission,chen2020nonabelian} and its large practical importance, unambiguous experimental signatures distinguishing between robust trivial ZESs in the helical regime and topological MBSs are still elusive.  

In this Research Letter we show how equilibrium phase-biased transport in long SNS junctions based on nanowires with SOC [see Fig.\,\ref{Fig1}(a)] can clearly distinguish between robust trivial ZESs and topological MBSs. In particular, we design a test of nonlocality, unique to MBSs, using supercurrents and critical currents by varying the lengths of the superconducting S regions. We demonstrate that these quantities are not affected by the S region lengths when trivial ZESs are present but exhibit strong changes when MBSs emerge. This different behavior occurs due to the Majorana nonlocality and thus provides a distinctive, yet experimentally easily accessible, way to separate topological MBSs from robust trivial ZESs by means of the Josephson effect. 


\emph{Nanowire junction}.---We consider long SNS junctions based on a single channel nanowire with Rashba SOC, using a continuum Hamiltonian given by
\begin{equation}
\label{Eq1}
H=\Big(\frac{p_{x}^2}{2m}-\mu\Big)\tau_{z}+\frac{\alpha_{\rm R}}{\hbar}p_{x}\sigma_{y}\tau_{z}+B\sigma_{x}\tau_{z}+\Delta(x)\sigma_{y}\tau_{y}\,,
\end{equation}
where $p_{x}=-i\hbar\partial_{x}$ is the momentum operator, $m$ the effective electron mass, $\mu(x)$ is the chemical potential dependent on position $x$, $\alpha_{\rm R}$ is the Rashba SOC strength, $B$ is the Zeeman field resulting from an external magnetic field, $\Delta(x)$ the induced $s$-wave superconducting order parameter, and $\sigma_{i}$ and $\tau_{i}$  are the $i$-Pauli matrices in spin and electron-hole spaces, respectively. In the normal state dispersion we find a gap of size $2B$ at zero momentum and centered around $\mu_{\rm N} =0$; see Fig.\,\ref{Fig1}(b). For $\mu_{\rm N}$ within this gap the system exhibits only two Fermi points (green circles), representing counterpropagating states with different spins \cite{PhysRevLett.90.256601}, thus generating a \textit{helical phase}.
Below, we show that this helical phase is particularly interesting as it allows the junction to host robust, but trivial ZESs.

In order to model finite systems,  we discretize the continuum Hamiltonian in Eq.\,(\ref{Eq1}) into a tight-binding lattice with spacing $a=10$\,nm  \cite{cayao2018andreev}. This discretized model is then divided into three regions of finite length; see Fig.\,\ref{Fig1}(a): The left and right S regions host a  finite superconducting order parameter $\Delta$, with a finite phase difference $\phi$, while the central N region is left with $\Delta=0$, giving rise to a finite SNS junction.  Each region has length $L_{\rm N, S}$ and finite chemical potential $\mu_{\rm N, S}$, tunable by means of voltage gates. We study SNS junctions with long N regions, $L_{\rm N}=2000$\,nm, but systematically vary the length of the S regions to differentiate between trivial ZESs and MBSs.
We further consider realistic system parameters: $\alpha_{\rm R}=40$\,meVnm and $\Delta=0.5$\,meV, which are in the range of experimental values reported for InSb and InAs nanowires, and Nb and Al superconductors \cite{lutchyn2018majorana}.


\begin{figure}[!ht]
	\centering
	\includegraphics[width=.49\textwidth]{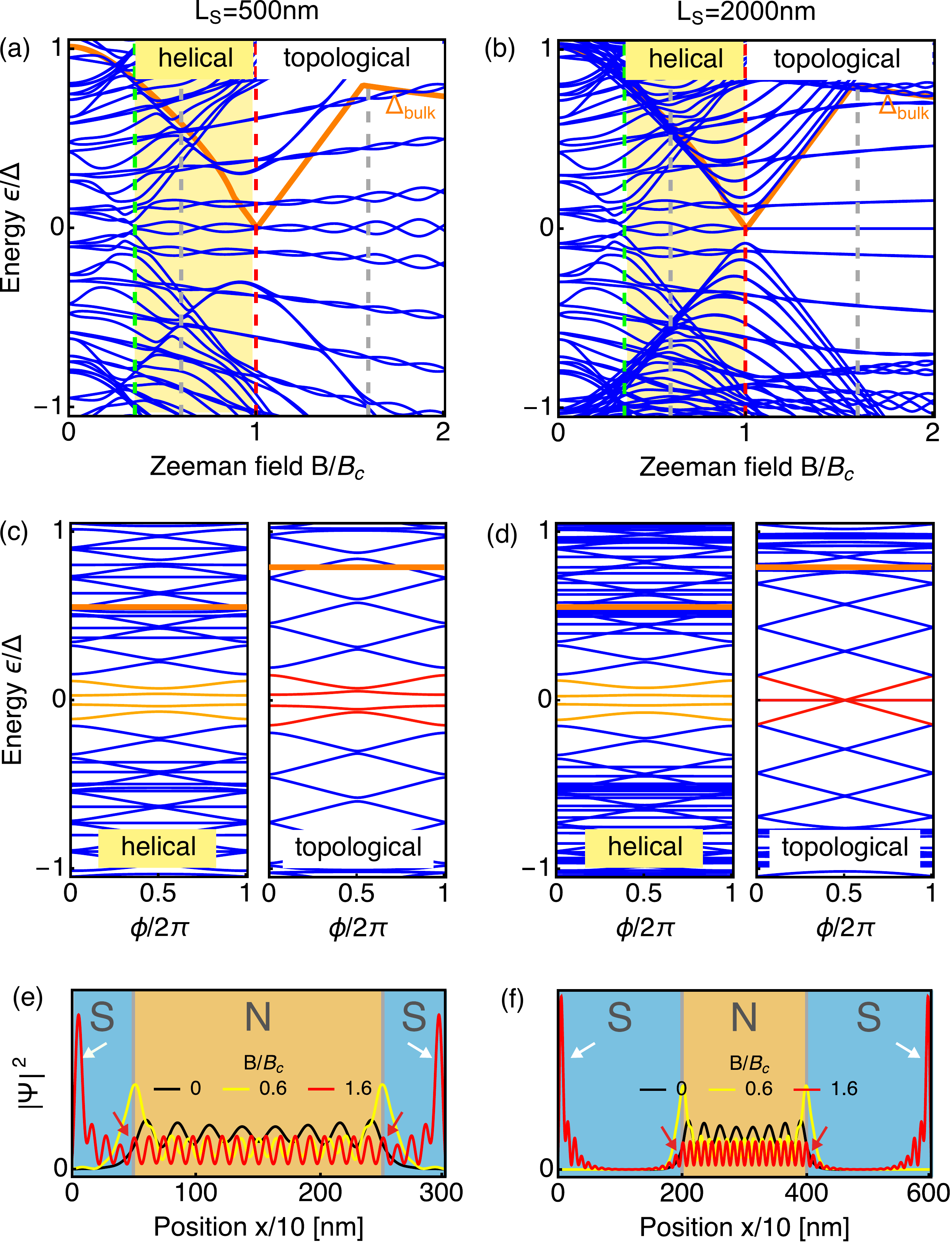}
	\caption{(a) and (b) Low-energy spectrum in a long SNS junction as a function of Zeeman field  $B$ at $\phi=0$ for $L_{\rm S}=500$\,nm (a) and $L_{\rm S}=2000$\,nm (b).  Vertical green and red dashed lines mark when the N region becomes helical (yellow region), $B=\mu_{\rm N}$, and the S regions become topological,  $B=B_{\rm c}$, respectively. The orange curve  corresponds to the minimum S bulk energy gap $\Delta_{\rm bulk}$.  (c)and (d) Phase-dependent low-energy spectrum taken at $B$ in the helical and topological regimes [vertical gray dashed lines in (a) and (b)] for both junctions with the lowest four levels color coded and the thick orange line indicating $\Delta_{\rm bulk}$ .  (e) and (f) Spatial dependence of the wavefunction amplitude $|\Psi|^{2}$ for the lowest four states [color coded in (c) and (d)] at $\phi=\pi$ at $B=0$ (black curve), in the helical phase (yellow curve), and the topological phase (red curve). Arrows indicate the outer (white arrows) and inner (red arrows) MBSs. 	
	 Here $\mu_{\rm S}=0.5$\,meV, $\mu_{\rm N}=0.25$\,meV.}
	\label{Fig2}
\end{figure}

\emph{Trivial ZESs and topological MBSs}.---To show the emergence of robust trivial ZESs and also contrast them with the topological MBSs, we first study the low-energy spectrum of long SNS junctions in Fig.\,\ref{Fig2}.  In Figs.\,\ref{Fig2}(a) and       \ref{Fig2}(b) we present the Zeeman dependent low-energy spectrum at $\phi=0$ for short and long S  regions, using $L_{\rm S}\leq2\ell_{\rm M}$ and $L_{\rm S}\geq2\ell_{\rm M}$, respectively, where $\ell_{\rm M}$ is the Majorana localization length.
At $B=0$, many low-energy levels appear within the S bulk energy gap $\Delta_{\rm bulk}$ as a result of having a long N region. These levels split at finite $B$, and the energy spectrum undergoes two transitions when $B$ is further increased, reflecting the nature of the N and S regions. 
First, at $B=\mu_{\rm N}$ (vertical dashed green line) the low-energy spectrum in the N region exhibits a gap inversion-like feature, which leads to energy levels around zero energy developing parity crossings that oscillate and, remarkably, persist as $B$ increases.  This regime occurs as a result of confinement and helicity of the N region in long SNS junctions. The developed ZESs are topologically trivial and emerge due to helical Fabry-Perot resonances in the N region \cite{PhysRevB.91.024514}. These resonances  reflect an instance of helical transport \cite{PhysRevB.90.235415,PhysRevB.91.024514,kammhuber2017conductance,doi:10.1021/acs.nanolett.8b01799,doi:10.1021/acs.nanolett.7b03854,doi:10.1063/5.0014148} present when the N region hosts only spin-polarized counterpropagating states in the helical phase \cite{PhysRevLett.90.256601}. We verify that these trivial ZESs start to form already for  $L_{\rm N}>250$\,nm, and are thus expected to be achieved in current experiments, see e.g.~Ref.~\cite{PhysRevLett.125.116803}. These trivial ZESs are remarkably robust against parameter changes, as long as the helical phase persists in  N.

Second, as $B$ further increases, the S regions enter into the topological phase at $B=B_{\rm c}$ [vertical red dashed line in Figs.\,\ref{Fig2}(a) and \ref{Fig2}(b)].  At this point, an energy gap inversion is expected in the spectrum, but due to finite lengths a complete gap closing is not present \cite{PhysRevB.93.245404}. We still indicate the bulk gap inversion obtained from Eq.\,(\ref{Eq1}) with $\Delta_{\rm bulk}$ (thick orange curve). 
In the topological phase, $B>B_{\rm c}$, two states emerge around zero energy and oscillate with $B$  due to the short length of S \cite{PhysRevB.86.085408,PhysRevB.86.180503,PhysRevB.86.220506,PhysRevB.87.024515}, particularly seen in Fig.\,\ref{Fig2}(a). These states are the \textit{outer MBSs}, located at the outer ends of the SNS junction. The would-be MBSs at the inner ends of the S regions hybridize strongly across the N region and are not present unless $\phi=\pi$ \cite{PhysRevB.96.205425,cayao2018andreev}. Note that these outer MBSs oscillate in a very similar fashion to the trivial ZESs in the helical phase, which complicates the distinction of states around zero energy. However, when the S regions become longer, the oscillations in the topological phase are reduced [see Fig.\,\ref{Fig2}(b), where $L_{\rm S}\geq 2\ell_{\rm M}$], while the behavior in the helical phase is essentially unresponsive to variations in $L_{\rm S}$.

To gain further evidence of the difference between trivial ZESs and topological MBSs, we analyze in Figs.\,\ref{Fig2}(c) and \ref{Fig2}(d) the low-energy phase-dependent spectrum at the $B$ fields in the helical and topological regimes marked by the  vertical gray dashed lines in Figs.\,\ref{Fig2}(a) and \ref{Fig2}(b). First, in both regimes we observe that the multiple energy levels emerging within the gap disperse linearly with $\phi$, an effect attributed to the long junction nature \cite{zagoskin,PhysRevLett.110.017003,PhysRevB.96.165415,cayao2018andreev}. Here, we distinguish the lowest four levels, shown in orange and red for the trivial and topological phases, respectively, from the higher levels also present within the bulk gap $\Delta_{\rm bulk}$,  as well as the quasicontinuum above $\Delta_{\rm bulk}$.  For short S regions  the phase-dependent low-energy spectra in the helical and topological phases are very similar. This indicates that it is extremely challenging  in this regime to identify whether the low-energy spectrum, and its associated ZESs, corresponds to a helical trivial or topological phase, as also argued in Ref.\,\cite{PhysRevB.91.024514}. 
It is, however, possible to find a prominent difference between the helical and topological phases by noting that the lowest four energy levels exhibit a distinct response to an increase in $L_{\rm S}$. This is illustrated in Fig.\,\ref{Fig2}(d), where we clearly see that, as a result of longer S regions, the splitting around zero energy is reduced in the topological phase, while the helical phase spectrum is almost unchanged.  This behavior can be understood by looking at the total wave function amplitude of the four lowest states in Figs.\,\ref{Fig2}(e) and \ref{Fig2}(f).  In the topological phase at $\phi=\pi$ the four lowest levels shown in red in Figs.\,\ref{Fig2}(c) and \ref{Fig2}d) represent four MBSs (red curves): The lowest two levels are the two outer MBSs (white arrows), while the next two levels are the two inner MBSs  present at $\phi=\pi$ only (red arrows) \cite{PhysRevB.86.140504,PhysRevLett.108.257001,PhysRevB.94.085409,PhysRevB.96.205425,cayao2018andreev}. We see that for short $L_{\rm S}\leq\ell_{\rm M}$, the MBS wave functions exhibit a very strong spatial overlap across the S region, which results in finite zero energy splitting at $\phi=\pi$. For long S with $L_{\rm S}\geq\ell_{\rm M}$, the spatial overlap is considerably reduced, and so is the energy splitting at $\phi=\pi$. In contrast, in the helical regime, the lowest states (yellow curves) are always mainly located at the SN and NS interfaces and do not exhibit a noticeable change with the length of S. Note that these helical states are different from the states at $B=0$, which are fully extended in the whole N region (black curves).

The suppression of the energy splitting at $\phi=\pi$ in the topological phase occurring due to the reduction in the spatial overlap between the MBS wavefunctions in the S regions can directly be interpreted as a result of the non-local nature of  MBSs. It is thus the MBS  nonlocality that causes the different behavior between the trivial helical and topological regimes. Therefore, by measuring the phase-dependent  spectrum in SNS junctions, e.g.~by  spectroscopic means as in Refs.~\cite{Lee:13,Doh:S05,nphys1811,Nilsson:NL12,Chang2013Tunneling,PhysRevLett.121.047001,PhysRevX.9.011010}, it should be possible to identify the non-local nature of the topological  MBSs and rule out any trivial ZES interpretation.

\begin{figure}[!t]
	\centering
	\includegraphics[width=.49\textwidth]{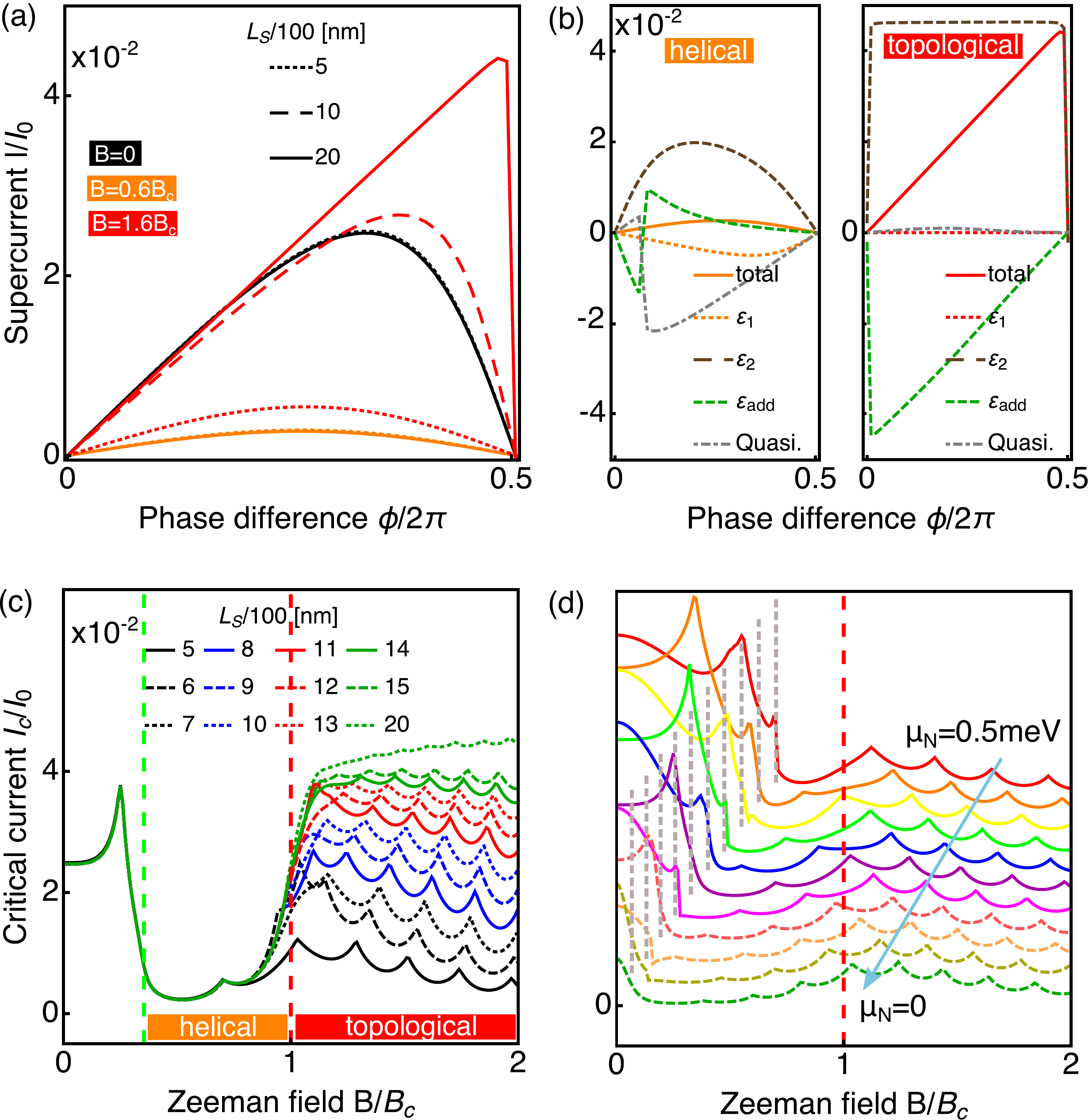}
	\caption{(a) Current-phase relationship for a long SNS junction in $B=0$ (black curves), helical (orange curves), and topological (red curves) phases and for different S region lengths $L_{\rm S}$. Here  $I_{0}=(e\Delta/\hbar)$. (b) Individual contributions to $I(\phi)$ from the first level $\varepsilon_{1}$, second level $\varepsilon_{2}$, additional levels below $\Delta_{\rm bulk}$ ($\varepsilon_{\rm add})$, and the quasicontinuum (Quasi; levels above $\Delta_{\rm bulk}$) for $L_{\rm S}=2000$\,nm. 
	(c) and (d) Critical current as a function of the Zeeman field $B$ for the same long junction as in (a) for different   $L_{\rm S}$ (c) and  different  $\mu_{\rm N}$ for $L_{\rm S}=500$\,nm (d). Curves in (d) are offset by a constant value for visualization. Vertical gray dashed lines in (d) mark the helical phase transition. The rest of  the parameters are as in Fig.\,\ref{Fig2}. 	}
	\label{Fig3}
\end{figure}

\emph{Current-phase relationship}.---In order to provide a more distinctive tool to distinguish trivial ZESs from topological MBSs than the Andreev spectrum discussed above, we study the phase dependent supercurrents. At zero temperature, the current is obtained as $I(\phi)=-(e/\hbar)\sum_{\varepsilon_n>0}[d\,\varepsilon_{n}(\phi)/d\phi]$, where $\varepsilon_{n}(\phi)$ are the phase dependent energy levels studied in the previous section. Since our system is of finite length, $I(\phi)$ automatically also includes the contribution of the discrete  phase-dependent   quasicontinuum \cite{cayao2018andreev}. The phase-dependent supercurrents $I(\phi)$ are presented in Fig.\,\ref{Fig3}(a) for different values of the S region length $L_{\rm S}$ and Zeeman field $B$. For a complete understanding, in Fig.\,\ref{Fig3}(b) we also plot the individual contribution of the energy levels to $I(\phi)$, separated into the first two levels $\varepsilon_{1,2}$ [ shown in orange and red in Figs.\,\ref{Fig2}(c) and \ref{Fig2}(d)];  additional levels below $\Delta_{\rm bulk}$, denoted here as $\varepsilon_{\rm add}$; and the quasicontinuum above $\Delta_{\rm bulk}$.

At $B=0$, $I(\phi)$ exhibits the usual sine-like form and its behavior does not change when increasing $L_{\rm S}$, as depicted by the black curves in  Fig.\,\ref{Fig3}(a).  At finite $B$ in the helical regime, the amplitude of $I(\phi)$ is reduced due to Zeeman depairing, but the overall sine-like profile of $I(\phi)$ and its independence on $L_{\rm S}$, remarkably, remain, see orange curves. The insensitivity of $I(\phi)$ to $L_{\rm S}$ for $B<B_{\rm c}$ is the usual situation in conventional trivial ballistic junctions \cite{Beenakker:92}, and here we  find that this also applies to junctions hosting trivial ZESs in the helical phase. Note that  most states contribute to the current profile, see Fig.\,\ref{Fig3}(b). 

The behavior is very different in the topological phase $B>B_{\rm c}$. Here, $I(\phi)$ instead exhibits a very strong dependence on $L_{\rm S}$, see  red curves in Fig.\,\ref{Fig3}(a). In junctions with short S regions $I(\phi)$ has a sine-like behavior and thus is very similar to the trivial helical regime, making them nearly impossible to distinguish. However, when  $L_{\rm S}$ increases, $I(\phi)$  acquires larger amplitudes in the topological phase  giving rise to very different current-phase relationships, in contrast to the superimposed curves in the helical regime; $I(\phi)$ also exhibits a notably more steep profile at $\phi=\pi$ that evolves into a sawtooth profile for very long S regions. This behavior can be understood by inspecting the phase-dependent low-energy spectrum in Figs.\,\ref{Fig2}(c) and  \ref{Fig2}(d), where we observe that the finite zero energy splitting at $\phi=\pi$ is considerably  reduced for large $L_{\rm S}$, stemming from the reduced spatial overlap of the outer and inner MBSs wavefunctions. This view is further supported by noting that it is the contribution from the second  level, $\varepsilon_{2}$, corresponding to the inner MBSs, that largely determines the fast sign change profile of  $I(\phi)$ at $\phi=\pi$, see Fig.\,\ref{Fig3}(b). In contrast, close to $\phi=0$, this contribution is largely canceled by the additional levels within the bulk gap. Contributions from the quasicontinuum and lowest energy level $\varepsilon_{1}$ (outer MBSs) are both negligibly small, in contrast to the helical regime. The fast sign change in $I(\phi)$ around $\phi=\pi$ thus results from the nonlocal nature of MBSs, which becomes a dominant feature with increasing length of the S region. Therefore, the $L_{\rm S}$ dependence  in the current-phase relationship offers a remarkably distinctive tool to identify  MBSs and trivial ZESs,   which clearly works even when the sawtooth profile smoothens.

\emph{Critical currents}.---An even simpler experimental tool to distinguish trivial ZESs from MBSs can be obtained by studying the critical current $I_{\rm c}$, which is the maximum supercurrent that can flow through the junction, $I_{\rm c}={\rm max}_{\phi}[I(\phi)]$ \cite{Beenakker:92,PhysRevB.96.205425}.  In Figs.\,\ref{Fig3}(c) and \ref{Fig3}(d) we present the Zeeman field dependent critical currents for different $L_{\rm S}$ and different $\mu_{\rm N}$, respectively.

As $B$ increases, the critical current undergoes a large change at $B=\mu_{\rm N}$ when the N region becomes helical [vertical green  dashed line  in Fig.\,\ref{Fig3}(c)]. As $B$ further increases within the helical regime, $I_{\rm c}$ reflects the oscillations with $B$ of the lowest energy levels, seen in Figs.\,\ref{Fig2}(a) and \ref{Fig2}(b), but the critical current is notably completely independent of variations in $L_{\rm S}$. At $B=B_{\rm c}$ the system enters the topological phase [vertical red dashed line in Figs.\,\ref{Fig3}(c) and \ref{Fig3}(d)], but this transition is largely unnoticed in $I_{\rm c}$, an effect due to contributions from additional levels in  the long N regions, in stark contrast to the case of  junctions with very short $L_{\rm N}$ \cite{SanJoseNJP:13,PhysRevLett.112.137001,PhysRevB.96.205425}.  In the topological phase, $I_{\rm c}$ exhibits notable oscillations as a function of $B$ for all shorter S lengths, due to the zero energy splitting at $\phi=\pi$. As $L_{\rm S}$ increases, $I_{\rm c}$ acquires larger values and the oscillations wash out, consistent with the vanishing of the zero-energy splitting at $\phi=\pi$. The strong dependence of $I_{\rm c}$ on $L_{\rm S}$ in the topological regime is a direct consequence of the nonlocal nature of MBSs,  which provides unique information even if the oscillations themselves are small. This stands in stark contrast to the insensitivity of  $I_{\rm c}$ to variations in $L_{\rm S}$ in the topologically trivial helical regime.

We further investigate the strong feature of $I_{\rm c}$ at the helical transition at $B=\mu_{\rm N}$  by  depleting carriers in  $N$, as presented in Fig.\,\ref{Fig3}(d) for a junction with short S regions (curves offset for visualization). When the chemical potentials in N and S are equal (uppermost red curve), $I_{\rm c}$ is strongly reduced at the helical transition $B=\mu_{\rm N}$ [vertical gray dashed  line]. As $\mu_{\rm N}$ is reduced, the helical transition moves to lower $B$, but it remains a prominent feature in $I_{\rm c}$. For all these different junctions there is no notable evidence of the topological phase transition [ vertical red dashed  line] in $I_{\rm c}$, which indicates that it is nearly impossible to know whether the system is in the topological phase from a single measurement. However, as we established in Fig.\,\ref{Fig3}(c), the topological phase is clearly detectable when probing the critical current as a function of length of the S regions. We further confirm that for longer S regions, only the oscillations arising from MBSs are washed out, while the signature of the helical transition  and the oscillations due to the helical ZESs remain robust  (not shown). 
 
We have verified that the above-reported findings are robust against  scalar disorder (of the order of $\mu_{\rm S}$), against reduction in  normal transmission at the SN interfaces, and also against finite temperatures (below the superconducting transition temperature). Moreover, we have also studied an alternative approach to test nonlocality   where we keep $L_{\rm S}$ fixed, with all our findings presented here remaining intact (see Supplemental Material \cite{SM}). Based on current developments, we also stress that the low-energy spectrum and supercurrents, including  critical currents, in SNS junctions, have been measured in junctions similar to the ones studied in this Research Letter \cite{tiira17,zuo17,Kjaergaard17,PhysRevLett.124.226801,PhysRevX.9.011010,PhysRevLett.125.116803}, and thus our findings provide an achievable route to distinguish trivial ZESs and topological MBSs.

In conclusion, we have demonstrated that robust trivial zero energy states and topological zero energy Majorana bound states, both emerging in long SNS junctions, can be clearly distinguished by means of equilibrium phase-biased transport in junctions with varying lengths of the S regions. These transport signatures are due to the unique nonlocality of the Majorana bound states and can thus be used to distinguish Majorana bound states from trivial zero energy states, even independent of the origin of the latter. 


We thank  J. C. Estrada Salda\~{n}a and E. Lee for insightful and motivating  discussions.  We acknowledge financial support from the Swedish Research Council (Vetenskapsr\aa det Grant No.~2018-03488), the Knut and Alice Wallenberg Foundation through the Wallenberg Academy Fellows program and the EU-COST Action CA-16218 NANCOHYBRI.

\bibliography{biblio}
\onecolumngrid
\appendix



\section*{Supplementary material}

# Supplemental Material for "Distinguishing trivial and topological zero energy states in long nanowire junctions"

Jorge Cayao and Annica M. Black-Schaffer

*Department of Physics and Astronomy, Uppsala University, Box 516, S-751 20 Uppsala, Sweden*

(Dated: July 3, 2021)

In this Supplemental Material we discuss an alternative approach to test non-locality in phase-biased supercurrents across long SNS junctions by maintaining fixed lengths of the superconducting regions S.

## ALTERNATIVE SCHEME TO TEST NON-LOCALITY IN PHASE-BIASED SUPERCURRENTS

In the main text of this work we present a protocol, based on phase-biased supercurrents, for distinguishing between trivial and topological zero-energy states, the latter also known as zero-energy Majorana bound states (MBSs), in long superconductor-normal-superconductor (SNS) junctions. This protocol relies on testing the spatial non-locality of the wave functions, a unique property of MBSs, while topologically trivial zero energy states do not exhibit this feature. More specifically, we show that a spatial non-locality test can be achieved by varying the length of the S regions, $L_{\rm S}$, which, albeit challenging, it is supported by recent advances in fabrication of similar junctions, see e.g. Ref. [1]. Interestingly, there are other alternative ways to test the spatial non-locality using fixed lengths of the S regions, $L_{\rm S}$, and we discuss below.

One alternative possibility is to compare a SNS junction with a magnetic field everywhere to another SNS junction with a magnetic field only partially applied in the S regions, see leftmost column in Fig. S1. For sufficiently large magnetic fields, the regions with an applied magnetic field become topological. While it is not simple to manipulate magnetic fields locally in these systems, recent experiments have shown promising possibilities using ferromagnets, see e.g. Refs. [2–4]. In this way we have two cases using the same fixed $L_{\rm S}$, but still with different lengths of the topological superconducting regions, $L_{\rm TS}$. In the first case, Case 1, $L_{\rm S} = L_{\rm TS}$, while in Case 2 we have $L_{\rm S} > L_{\rm TS}$. As a consequence, since MBSs appear at the ends of the topological regions, in case 2 the MBSs develop a larger spatial wave function overlap across the junction. Just as in the situation discussed in the main text, this gives rise to a finite zero-energy energy splitting at $\phi = \pi$ in the low-energy spectrum due to MBSs.

### Low-energy spectrum

To visualize the discussion of the previous paragraph, in Fig. S1 we present the Zeeman- (a,b,e,f) and phase-dependent (c,d,g,h) low-energy spectrum for Case 1 (upper row) and Case 2 (lower row). The length of the region with magnetic field in S in Case 2 is here considered to be $L_{\rm S}/2$. We observe that the junction hosts topologically trivial zero-energy states, for a magnetic field large enough to make the N region helical (green vertical line lines), in agreement with the discussion in the main text. This occurs both at $\phi = 0$ and $\phi = \pi$, and, as discussed in the main text, the behavior of these trivial zero-energy states is the same for Case 1 (a,b) and Case 2 (e,f). In contrast, in the topological phase, obtained past the red vertical dashed line, we clearly notice that the lowest energy levels behave differently, with larger energy splitting in Case 2 compared to Case 1. This is even more evident when analyzing the phase-dependent low-energy spectrum for these two cases at Zeeman fields in the trivial helical and topological phases (marked by vertical grey dashed lines). We note that the lowest four energy levels in the topological phase (the MBSs) behave differently in the two cases (red levels in (d,h)), while the trivial states (orange in (c,g)) are unchanged between the two cases. This is because in Case 1 the length of the topological region is larger than in Case 2 and thus the zero-energy energy splitting in the topological phase in Case 1 at $\phi = \pi$ is much smaller than for Case 2. As discussed in the main text, the zero-energy splitting at $\phi = \pi$ occurs due to the spatial overlap between MBS wavefunctions. As a result, the different behavior of the lowest four levels in the topological phase reflects the spatial-non locality of MBSs, in the same way as discussed in the main text and there showed in Fig. 2.

### Current-phase relationship

To further support the previous discussion, we present in Fig. S2 the phase dependent supercurrents $I(\phi)$ for two values of the Zeeman field, in the trivial and topological regimes, marked by vertical grey dashed lines in Fig. S1. In both Case 1 and Case 2, the supercurrent is $2\pi$-periodic and develops a sine-like behavior but also exhibit interesting different properties. In the trivial helical regime, $I(\phi)$ is the same for Cases 1 and 2, leading to two superimposed supercurrents in Fig. S2, depicted by orange solid and dashed curves. Interestingly, in the topological phase, $I(\phi)$ instead senses a clear difference

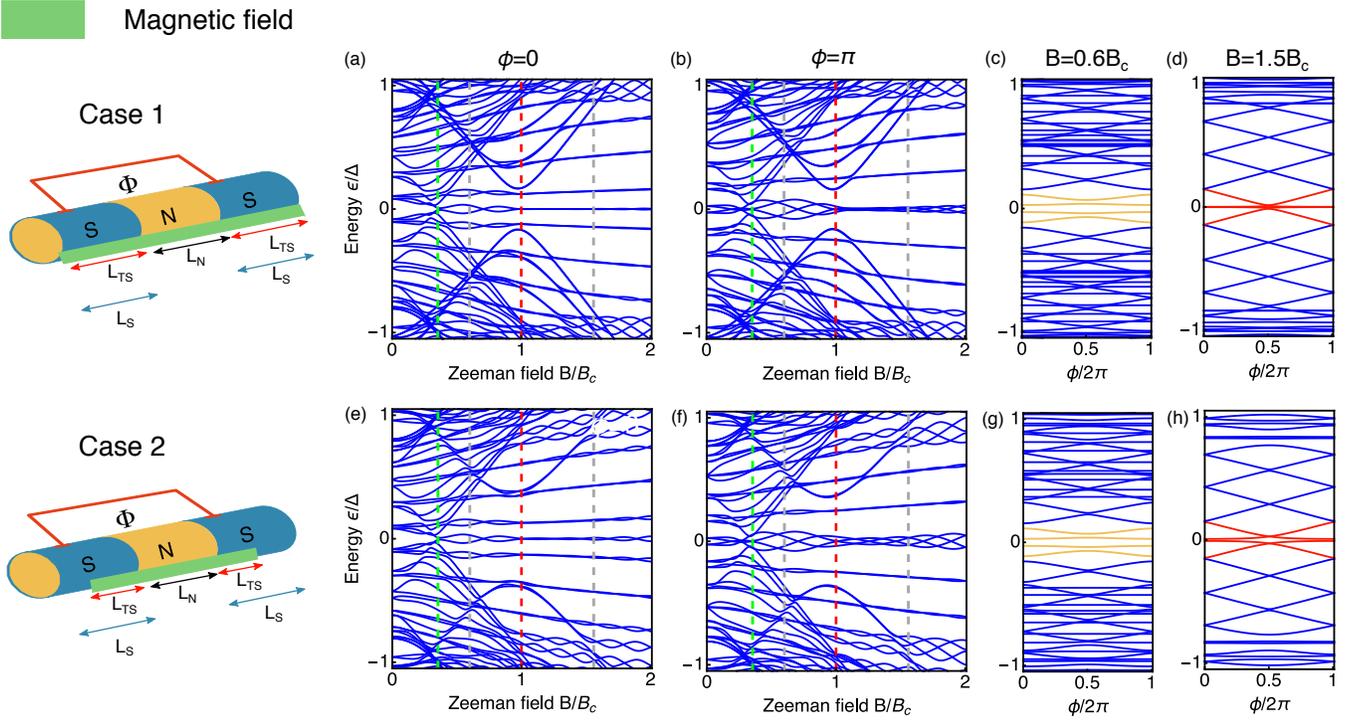

FIG. S1. (Color online.) Alternative scheme to test non-locality in a long SNS junction using fixed lengths of the S region: a magnetic field (green region) is applied to the whole nanowire (Case 1, leftmost top panel), or a magnetic field is applied only partially (Case 2, leftmost bottom panel). While in Case 1 the length of S is the same as the topological region, $L_S = L_{TS}$, in Case 2 the topological sector is smaller, $L_S > L_{TS}$. Panels (a,b) and (e,f) correspond to the Zeeman-dependent low-energy spectrum for Case 1 and Case 2, respectively. Here, vertical green and red dashed lines mark where the N region becomes helical, $B = \mu_N$, and the S regions becomes topological, $B = B_c$, respectively. Panels (c,d) and (g,h) correspond to the phase-dependent low-energy spectrum for Case 1 and Case 2, respectively, taken at specific $B$-values in the helical and topological regimes (vertical gray dashed lines in (a,b) and (e,f)). Note that because the magnetic field is restricted to a shorter region in Case 2, the lowest energy levels in the topological phase are changed compared to Case 1, a result of the spatial non-locality of the MBSs. Parameters: $L_N = 2000$ nm, $L_S = 1000$ nm, $\mu_S = 0.5$ meV, $\mu_N = 0.25$ meV, $\Delta = 0.5$ meV, $\alpha_R = 40$ meVnm.

between the two cases, resulting in supercurrents with notably different amplitudes for the two cases (red dashed and solid lines), for the parameters used here, we find a more than 50% increase. Note that for Case 1, where $L_S = L_{TS}$, the supercurrent develops both larger amplitudes and a faster sign change at $\phi = \pi$ than in Case 2, where $L_S > L_{TS}$. Notably though, the fast sign change of $I(\phi)$ in the topological phase at $\phi = \pi$ does not really develop into a sawtooth profile as found in the main text. Still, it is very clear that the difference in $I(\phi)$ between Cases 1 and 2 allow for unambiguously distinguishing between the trivial and topological regimes.

The very different behavior of $I(\phi)$ in the trivial (orange curves in Fig. S2) and topological (red curves) regimes is a direct consequence of the behavior of the phase-dependent low-energy spectrum presented in Fig. S1(c,d,g,h). In fact, it is the zero-energy splitting at $\phi = \pi$ in the topological phase that is causing this difference in the supercurrent, in a very similar way as discussed in the main text where instead we vary the lengths of S. This, in turn, reflects the spatial non-locality of the MBSs, thus connecting supercurrent measurements with

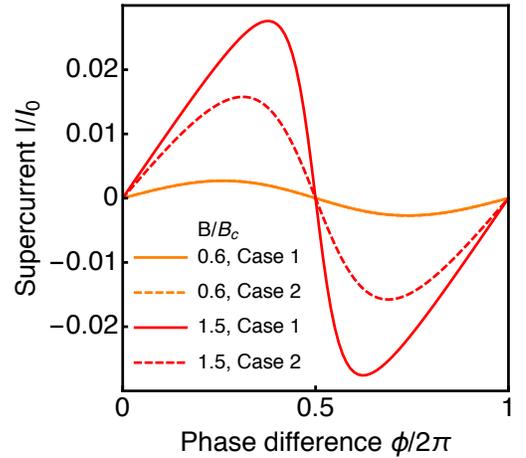

FIG. S2. Current-phase relationship for a long SNS junction at fixed length of S in the helical (orange) and topological (red) phases for Case 1 and Case 2 discussed in Fig. S1. Here $I_0 = (e\Delta/\hbar)$, with all parameters as in Fig. S1.

S3

the non-locality of MBSs, in exactly the same way as for the junctions discussed in the main text. Therefore, we conclude that a test of non-locality can be also achieved at fixed lengths of S, $L_{\rm S}$, by allowing the magnetic field to act on different parts of the S regions, and in that way achieve topological sectors of different lengths.

We here point out that the supercurrents presented in Fig. S2 do not exhibit a sawtooth profile at $\phi=\pi$ because the topological sectors are not sufficiently long as the ones considered in the main text, see e.g. solid red curve in Fig. 3(a) in main text. In spite of this, the different response of the supercurrents to the test of non-locality in the trivial and topological phases is a strong and highly measurable property. This thus allows us to conclude that the test of non-locality in phase-biased supercurrents that we propose can be used even when the current-phase relationship do not exhibit any sawtooth profile, which is a very likely situation in experiments due to e.g. finite temperature effects. As a consequence, the test of non-locality we propose in this work is a powerful, yet simple, tool to distinguish zero-energy states in the trivial and topological regimes.

Following the implementation discussed in this Supplemental Material, it is also possible to test the MBS non-locality at fixed $L_{\rm S}$ and for an homogeneous magnetic field but by applying different external voltage gates to the S regions, such that only part of S becomes topological. This way we yet again achieve systems with different topological region lengths and can thus perform tests of non-locality in phase-biased supercurrents, as we propose here.



\end{document}